\documentclass[preprint,review,12pt]{elsarticle}
\usepackage{amssymb}
\usepackage[numbers]{natbib}
\usepackage{lineno}
\usepackage{amsmath,environ}
\usepackage[utf8]{inputenc}
\usepackage[T1]{fontenc}
\usepackage[utf8]{inputenc}
\usepackage{tabularx,ragged2e,booktabs,caption}
\usepackage{booktabs}
\usepackage{longtable}
\usepackage[nolists]{endfloat}
\usepackage[font=small,labelfont=bf]{caption}
\setcitestyle{round}
\biboptions{authoryear, round}
\usepackage[margins]{trackchanges}

\setcitestyle{square,aysep={}}
\addeditor{RW}
\addeditor{HT}
\journal{Marine Geology}
\begin{document}
\begin{frontmatter}

\title{A Model for TSUnami FLow INversion from Deposits (TSUFLIND)}
\author[geovt]{Hui Tang\corref{cor1}}
\ead{tanghui@vt.edu}
\author[geovt]{Robert Weiss}
\cortext[cor1]{Corresponding author}
\address[geovt]{Department of Geoscience, Virginia Polytechnic State University, United States}

\begin{abstract}

-

\end{abstract}

\end{frontmatter}

\section{Introduction}
\label{S:1}
The tsunami events that occurred over the last decades have caused an increase in public awareness and resulted in more research on the tsunami wave. Tsunami deposits play an important role not only in tsunami hazard assessments, but also in interpreting tsunami hydraulics \citep{hutchinson1997,Moore2007336,Jaffe2007347}. To draw any useful quantitative conclusions from tsunami deposits, the information from deposits about the causative tsunami needs to be extracted either by comparing parameters from the deposits with results from forward models \citep[see][]{Bourgeois1988,martin2008} or by inversion models directly
\citep[see][]{Nott1997193, Noormets200441, Jaffe2007347, Moore2007336, sousby2007, Smith2007362, benner2010,   Nandasena2013163}.

Tsunami inversion models attempt to link the basic information of the tsunami deposits with the overland flow characteristics. There are three prominent inversion models: Moore's advection model \citep{Moore2007336}, Soulsby's model \citep{sousby2007} and TsuSedMod model \citep{Jaffe2007347}. It should be noted that all these models are based on different basic assumptions and employ different information from the deposits. For example, Moore's advection model estimates tsunami flow magnitude by determining the combination of flow velocity and depth to move the largest grain from the sediment source to the deposition area \citep{Moore2007336}. In this paper, we present a joint inversion framework (TSUFLIND), which combines these three models. TSUFLIND does not only couple all these three inversion models, but also contains a new method to calculate deposit characteristics \citep{HUI2014}. It also uses the calculated flow depth from Soulsby's model to estimate a representative offshore tsunami wave amplitude. 

\section{Theoretical Background}
\label{S:2}
\subsection{Inversion Models Employed}
As mentioned above, there are three prominent tsunami deposition inversion models that will be used: Moore's advection model, Soulsby's model and TsuSedMod model.

\begin{figure}[h!]
  \centering
    \includegraphics[width=1\textwidth]{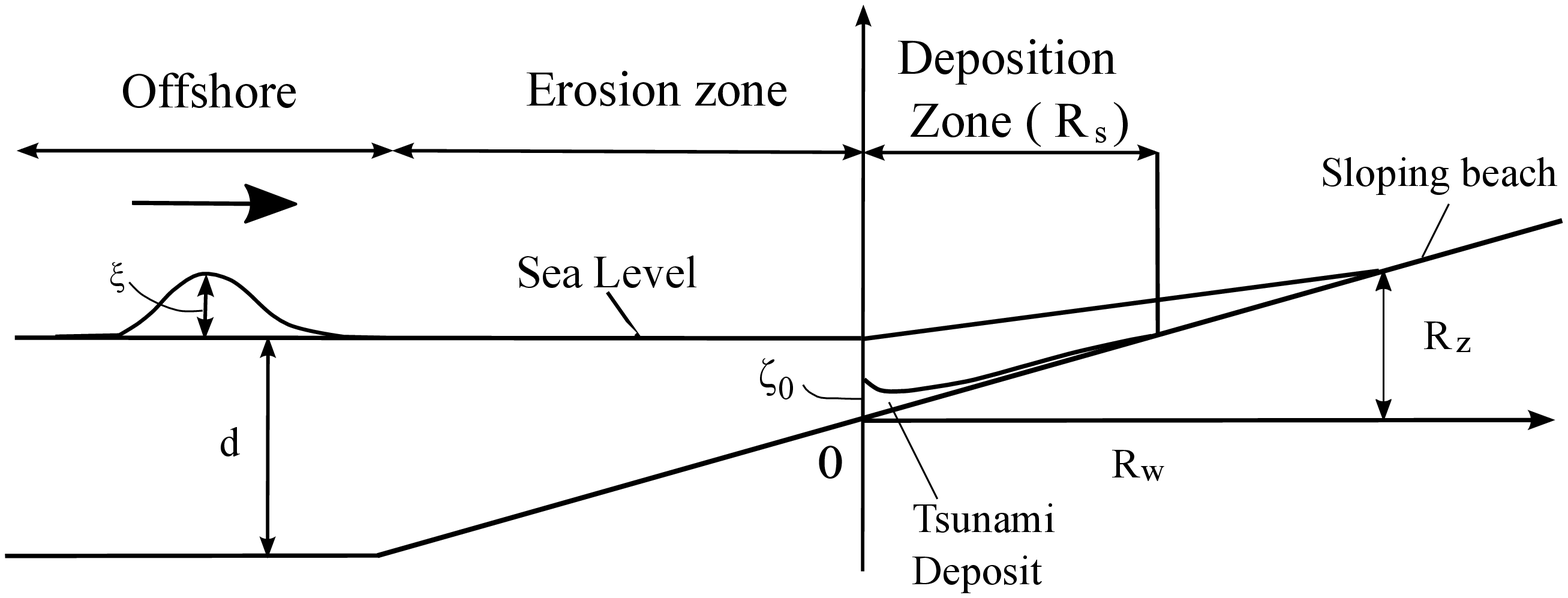}
  \caption{Conceptual model of TSUFLIND with definition of the terminology used later in the paper. For more symbols used in this paper see Appendix I.}
\end{figure}

\paragraph{(a) Moore's model:}
\citet{Moore2007336} assumes that some grains in the sediment source do not move because the tsunami flow is not strong enough. Furthermore, it is assumed that most grains are transported in suspension. Based on these assumptions, the shear velocity is determined for the largest grain in the tsunami deposits. The law of the wall can be employed to find the shear stress, which is necessary to move the largest grain to get a flow velocity $U$. The following equation is used to determine deposition. 

\begin{equation}
\label{eq:1}
\frac{h}{w_{s}}=t=\frac{l}{U}
\end{equation}
in which $w_{s}$ is the settling velocity of the sediment grain. $h$ is the water depth, $l$ represents the horizontal distance a grain travels to be deposited. Because of the horizontal transport, this model is also referred as an advection model. 

This model was applied to deposits formed by the 1929 Grand Banks tsunami, Newfoundland, Canada \citep{Moore2007336}. In this application, it was estimated that the average flow depth was 2.5 to 2.8 m, and the flow speed was 1.9 to 2.2 $\text{ms}^{-1}$, which are the minima\citep{Moore2007336}.

\paragraph{(b) Soulsby's model:}
Soulsby's model assumes that the water depth increases linearly between $0$ and $\gamma T$ and decreases from $\gamma T$ to $T$ for any given locations. $T$ is the inundation time and $\gamma$ is a constant related to run-up time, which is between $0$ and $1$. $H=H_0+\Delta h$ is the maximum flow depth at a given location during tsunami inundation and decreases toward the inundation limit, $H_0$ denotes the maximum water depth at the shoreline, $\Delta h$ denotes the depth increment due to tsunami:

\begin{equation}
\label{eq:2}
\Delta h=\frac{l(R_{z}-H_0)}{mR_{z}}-\frac{l}{m}
\end{equation}
where $m$ is the slope and $R_{z}$ represents the vertical inundation limit. The thickness of the deposit for grain size $i$ at the shoreline:

\begin{equation}
\label{eq:3}
\zeta_{0}^{(i)}=\frac{C_{0}^{(i)}w_{s}^{(i)}T_d}{(1-p)\rho_{s}(1+\alpha^{(i)})}(1+\alpha^{(i)}\gamma)
\end{equation}
where  $\alpha^{(i)} = \frac{w_{s}^{(i)}T_d}{H_0}$, $w_{s}^{(i)}$ denotes the settling velocity for grain size $i$, $T_d=(1-\gamma)T$ is the deposition time. $C_{0}^{(i)}$ is the depth averaged sediment concentration for grain size $i$ and $p$ is the porosity. The sediment thickness for grain size $i$ linearly decreases with distance from the shoreline:
\begin{eqnarray}
\label{eq:4}
\zeta^{(i)}(x)=\left\{ \begin{array}{rl}
\zeta_{0}^{(i)}(1-\frac{x}{R_{s}^{(i)}}) & \textrm{ $x < R_{s}^{(i)}$}\\
0 & \textrm{ $x \geq R_{s}^{(i)}$}\\
\end{array}\right.
\end{eqnarray}
where $R_{s}^{(i)}$ is the distance between sediment extend and the shoreline for grain size $i$ \citep{sousby2007}.

\paragraph{(c) TsuSedMod:}
\citet{Jaffe2007347} developed an inversion model based on sediment deposited from suspension. There are several basic assumptions in TsuSedMod: (1) sediment is transported in suspension and deposited when steady and uniform tsunami flow slows down; (2) suspended sediment concentration is distributed in an equilibrium profile; (3) there is no erosion caused by return flow. The model iteratively adjusts the sediment source and the shear velocity to match the sediment grain-size distributions and thickness of suspension-grading sediment layers \citep{Jaffe201123,Jaffe201290}. For the grain size $i$, the sediment thickness $\Delta \eta^{(i)}$ is given by:

\begin{equation}
\label{eq:5}
\ \Delta \eta^{(i)}=\frac{1}{(1-p)}\int_{0}^{H(x)}{C^{(i)}(z)dz}
\end{equation}
where $C^{(i)}(z)$ is the sediment concentration profile of grain size $i$. After determining the shear velocity, the flow speed profile is calculated by :

\begin{equation}
\label{eq:6}
\ U(z)=\int_{z_0}^{z}{\frac{u_*^{2}}{K(z)}dz}
\end{equation}
where $z_o$ is the bottom roughness from \citet{MacWilliams2005} and $K(z)$ is the eddy viscosity profile from \citet{gelfenbaum1986}. 

The TsuSedMod model has been applied to four modern tsunami cases \citep{Jaffe2007347,Spiske201029,Jaffe201123,Jaffe201290} and two paleotsunami cases \citep{witter2012,Spiske201331}. For the 2009 tsunami near Satitoa, Samoa, the flow speed estimated from TsuSedMod at three locations (100, 170 and 240 meters inland) were 3.6 to 3.8 $\text{ms}^{-1}$ (bottom layer/earlier wave) and 4.1 to 4.4 $\text{ms}^{-1}$ (top layer/later wave). These results are consistent with the 3 to 8 $\text{ms}^{-1}$ flow speed from the boulder transport inverse model \citep{Jaffe201123}. 

\subsection{Sedimentation model}
The method used to calculate the sediment concentration of the sediment source in TSUFLIND is similar to the one presented in \citet{Madsen19931303}. The grain-size distribution of the sediment source is characterized by $D_{50}$, the largest grain and the smallest grain.

When the entire tsunami deposit at a given location is considered, resuspension sediment flux can be neglected and Soulsby$'$s model is applied. However, if the individual layer in the tsunami deposit is considered, intense turbulent mixing cannot be ignored. Therefore resuspension has to be taken into account. The generation of each individual portion of the tsunami sediment based on flow condition is the fundamental part to reconstruct tsunami deposits. For the entire deposit, the basic process is to calculate sediment thickness $\zeta^{(i)}(x)$ for each grain size at each point along the slope by using Eq (\ref{eq:3}) and (\ref{eq:4}) from Soulsby's model. We assume that the depth averaged sediment concentration $C_0$ in Eq (\ref{eq:3}) is the reference sediment concentration $C_{r}$ here. The reference concentration is calculated for a given flow condition with \citet{Madsen19931303}:

\begin{equation}
\label{eq:7}
C_{r}^{(i)}=\frac{\beta_{0}(1-p)f^{(i)}S^{(i)}}{1+\beta_{0}S^{(i)}}
\end{equation}
where $\beta_{0}$ is the resuspension coefficient, $f^{(i)}$ is a fraction of the sediment of grain size $i$. $S^{(i)}$ is the normalized excess shear stress given by 

\begin{eqnarray}
\label{eq:8}
S^{(i)}=\left\{ \begin{array}{rl}
\frac{\tau_{b}-\tau_{cr}^{(i)}}{\tau_{cr}^{(i)}} &\mbox{$\tau_{b} > \tau_{cr}^{(i)}$} \\
0 &\mbox{ $\tau_{b} \leq \tau_{cr}^{(i)}$}
\end{array}\right.
\end{eqnarray}
where $\tau_{b}$ is the bed shear stress and $\tau_{cr}^{(i)}$ is the critical shear stress of the initial sediment movement for grain size $i$ \citep{Madsen19931303}.  

For a given location $x$, the grain-size distribution for the entire tsunami deposit is given by:

\begin{equation}
\label{eq:9}
f^{(i)}=\frac{\zeta^{(i)}(x)}{\sum_{i=0}^{N}\zeta^{(i)}(x)}; i=1,2,3,\ldots,N
\end{equation}
where $f^{(i)}$ is the percentage of grain size $i$ in the entire sediment, $\zeta^{(i)}(x)$ is sediment thickness of grain size $i$ and $\sum_{i=0}^{N}\zeta^{(i)}(x)$ is total deposit thickness for all grain sizes. $N$ is the number of grain size classes.

The tsunami deposit characteristics are reconstructed by matching sediment thickness and grain-size distribution with field data. In order to reconstruct deposit details, the sediment concentration cannot be depth averaged and is described as a Rouse-type suspended sediment concentration profile. In this framework, we use the method from \citet{Jaffe2007347} to calculate the suspended sediment concentration profile. It is efficient to reconstruct the deposit by calculating times of deposition. The deposition time of suspended sediment is calculated by:

\begin{equation}
\label{eq:10}
t^{(i)}_{j}=\frac{z_{j}}{w_{s}^{(i)}}
\end{equation}
in which $t^{(i)}_{j}$ is the deposition time for grain size $i$ sediment at elevation $z_{j}$. The amount of sediment settling in each grain size class for a given elevation is tracked as sediment thickness increment:

\begin{equation}
\label{eq:11}
\zeta^{(i)}_{j}=\frac{C^{(i)}_{j}}{1-p}
\end{equation}
$C^{(i)}_j$ is from the suspended sediment profile \citep{Jaffe2007347}, $\zeta^{(i)}_{j}$ is the thickness of the sediment increment of the same grain size $i$ at elevation $z_{j}$ and deposited at time $t^{(i)}_{j}$. The deposition time and  corresponding sediment thickness increment are ordered from shortest to longest. If there are multiple layers in the tsunami sediment, we can compute the grain-size distribution for each layer separately based on the depositional temporal order of the sediment thickness increments by:

\begin{equation}
\label{eq:12}
f_{k}^{(i)}=\frac{\sum_{j=0}^{M}{\zeta^{(i)}_{j}}}{\sum_{i=0}^{N}{\Big (\sum_{j=0}^{M}{\zeta^{(i)}_{j}}}\Big )}; i=1,2,3,\ldots,N;j=1,2,3,\ldots,M
\end{equation}
where $f_{k}^{(i)}$ is the sediment fraction of grain size $i$ in layer $k$.  $\sum_{j=0}^{M}{\zeta^{(i)}_{j}}$ is total sediment thickness with the same grain size $i$ in sediment layer $k$. Index $j$ is used to mark the original location of sediment in the water column. $\sum_{i=0}^{N}{\Big (\sum_{j=0}^{M}{\zeta^{(i)}_{j}}}\Big )$ is the total thickness of this sediment layer which contains all grain size classes. In TSUFLIND, the calculation of tsunami flow condition will use the same method as TsuSedMod model \citep{Jaffe2007347}.

\subsection{Result Evaluation}
We employ the second norm to quantify the error between model and observed results as a control of the iterative procedure. The second norm of error for layer $k$ is given by:

\begin{equation}
\label{eq:13}
{L_{k}}=\sqrt{\frac{{\sum_{i=1}^{N}{\big (f^{(i)}_{m}-f^{(i)}_{o}\big)^2}}}{N}}
\end{equation}
$f^{(i)}_{m}$ and $f^{(i)}_{o}$ are the modeled and observed percentages for each grain size class $i$. With the help of $L_{k}$, we compute the average second norm value for a location with:

\begin{equation}
\label{eq:14}
\overline{L}=\frac{1}{K}\sum_{k=1}^{K}{L_{k}}
\end{equation}
We define $\overline{L} \leq 5\%$ as a good simulation. For the tsunami sediment thickness simulation, we employ the same process. The second norm value of error for thickness between the model result and the field observation is given by:

\begin{equation}
\label{eq:15}
{L_{th}}=\sqrt{\frac{\sum_{j=1}^{Q}{\Big (\frac{th_{m}-th_{f}}{th_{f}}\cdot 100\% \Big)^2}}{Q}}
\end{equation}
where $th_{m}$ and $th_{f}$ are the modeled and observed thicknesses for each sample location, $Q$ is the number of sample locations. As there is only a limited number of tsunami deposit samples for the test case applied here, we use 10\% as the threshold value.

\subsection{Offshore Wave Characteristics and Flooding}

In order to estimate a representative offshore tsunami amplitude, we relate the water volume calculated from Sousby's model at maximum inundation with the volume calculated by numerically solving the shallow water equation. We carry out a parameter study by varying the slope ($m$) and the offshore wave amplitude ($\xi$). For more details about the parameter study and employed numerical model, we refer to Appendix II. The water depth computed from Soulsby's model is used to calculate the volume of the inundation water. With the help of numerical simulations (Appendix II), we derived the following formulation:

\begin{equation}
\label{eq:16}
\begin{split}
\xi= & \lambda _1+\lambda _2\cdot V+\lambda _3\cdot m+\lambda _4\cdot V^2+\lambda _5\cdot m\cdot V+\lambda _6\cdot m^2 \\
& +\lambda _7\cdot V^3+\lambda _8\cdot V^2\cdot m+\lambda _9\cdot V\cdot m^2+\lambda _{10}\cdot m^3 
\end{split}
\end{equation}
Where $\xi$ is offshore wave height, $V$ is the water volume that covers the land at maximum inundation, $m$ is the slope of beach profile. These constants $\lambda$ in Eq 16 are $\lambda _1=5.06$, $\lambda _2=2.93$, $\lambda _3=-0.28$, $\lambda _4=0.51$, $\lambda _5=-3.04$, $\lambda _6=0.0014$, $\lambda _7=0.027$, $\lambda _8=-0.011$, $\lambda _9=0.051$, $\lambda _{10}=0.053$.

\begin{figure}[h!]
  \centering
    \includegraphics[width=0.9\textwidth]{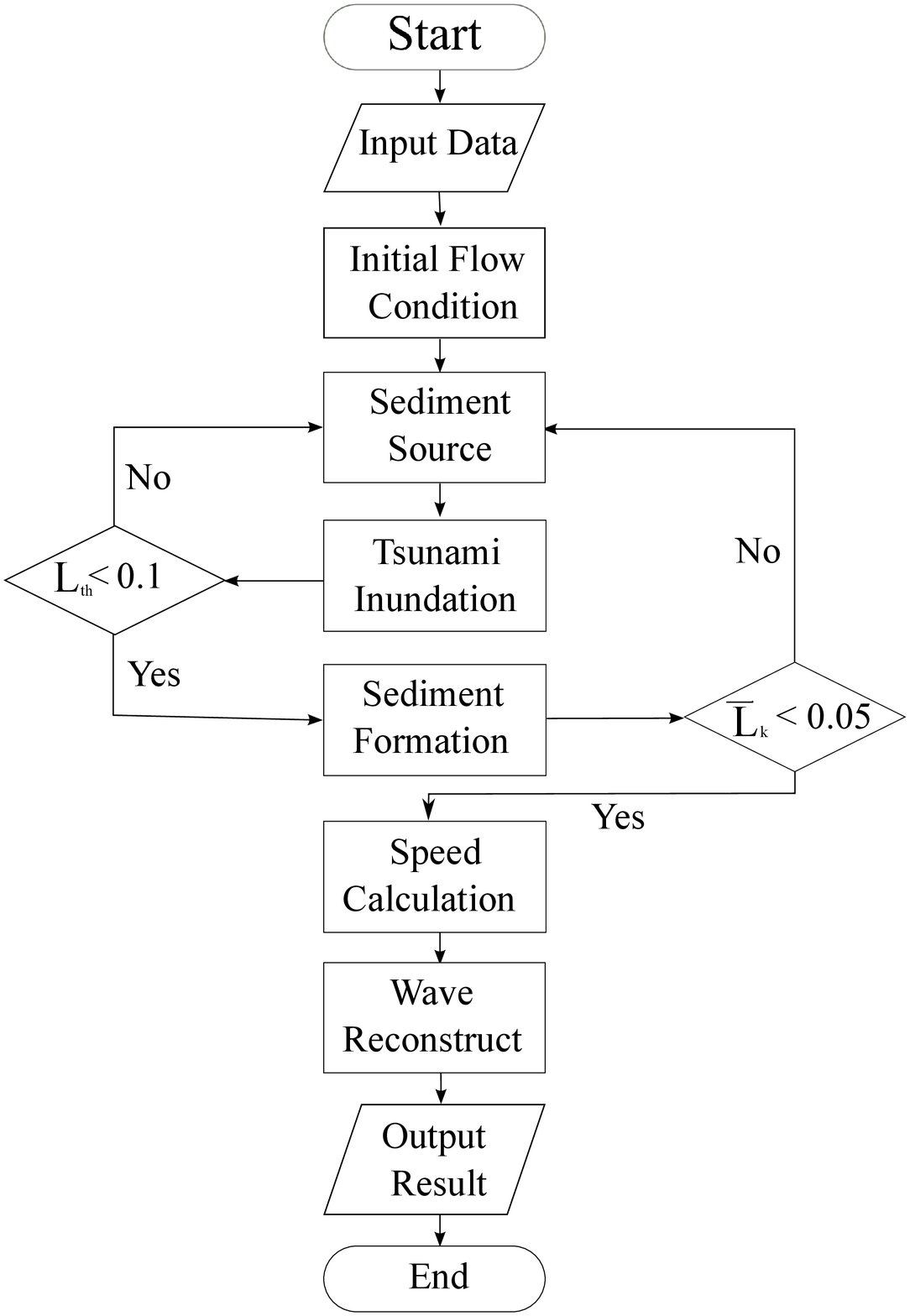}
  \caption{Flowchart for TSUFLIND's iterative scheme to simulate tsunami deposit and estimate tsunami flow condition. }
\end{figure}
\subsection{Inversion Framework and Coupling}

We use the information from all three models as different components in this joint inversion framework. The steady flow condition that is presented in all models, is also presented in TSUFLIND and represents the most simplifying assumption. The inputs to TSUFLIND are the sediment characteristics for different sampling locations along a slope. However, it should be noted that the inversion of the flow conditions is carried out for each sample location individually. TSUFLIND uses components from Moore model, Soulsby's model and TsuSedMod  model to adjust the sediment source grain-size distribution, the sediment source concentration and the average flow velocity to simulate tsunami sediment thickness and grain-size distribution along the slope in the deposition zone. If needed, the representative offshore wave amplitude can be computed. Figure 2 depicts the flowchart outlining how the joint inversion model works. 

The information needed for a successful inversion includes the grain-size distribution, sediment thickness as well as the information of the slope along which the tsunami sediments were sampled. In the inversion framework, Moore’s advection model is employed to calculate the initial flow speed. Because the Moore's model uses the actual data from the measured sediment distribution, it reduces the number of iterations significantly. The reservoir of sediments in the water column is calculated by following \citet{Madsen19931303}, and it is assumed that all grain-size distributions can be described with log-normal distributions. The iteration begins with computing the inundation with the help of Soulsby’s model, and the initial estimates of the flow conditions are from the Moore's advection model. The results of this step are the local flow depth and the entire sediment thickness at each sample location. Our sediment formation module calculates the characteristics of the deposited sediments. The iterations are controlled by the norm of error between the simulated and observed deposits and stop after the predefined threshold is met. As the model outputs, we can estimate the flow speed, depth and Froude number along the slope. If needed, a range of offshore reference wave amplitudes can also be computed.

\section{Application and Example}
\label{S:3}

\begin{figure}[h!]
  \centering
    \includegraphics[width=1.0\textwidth]{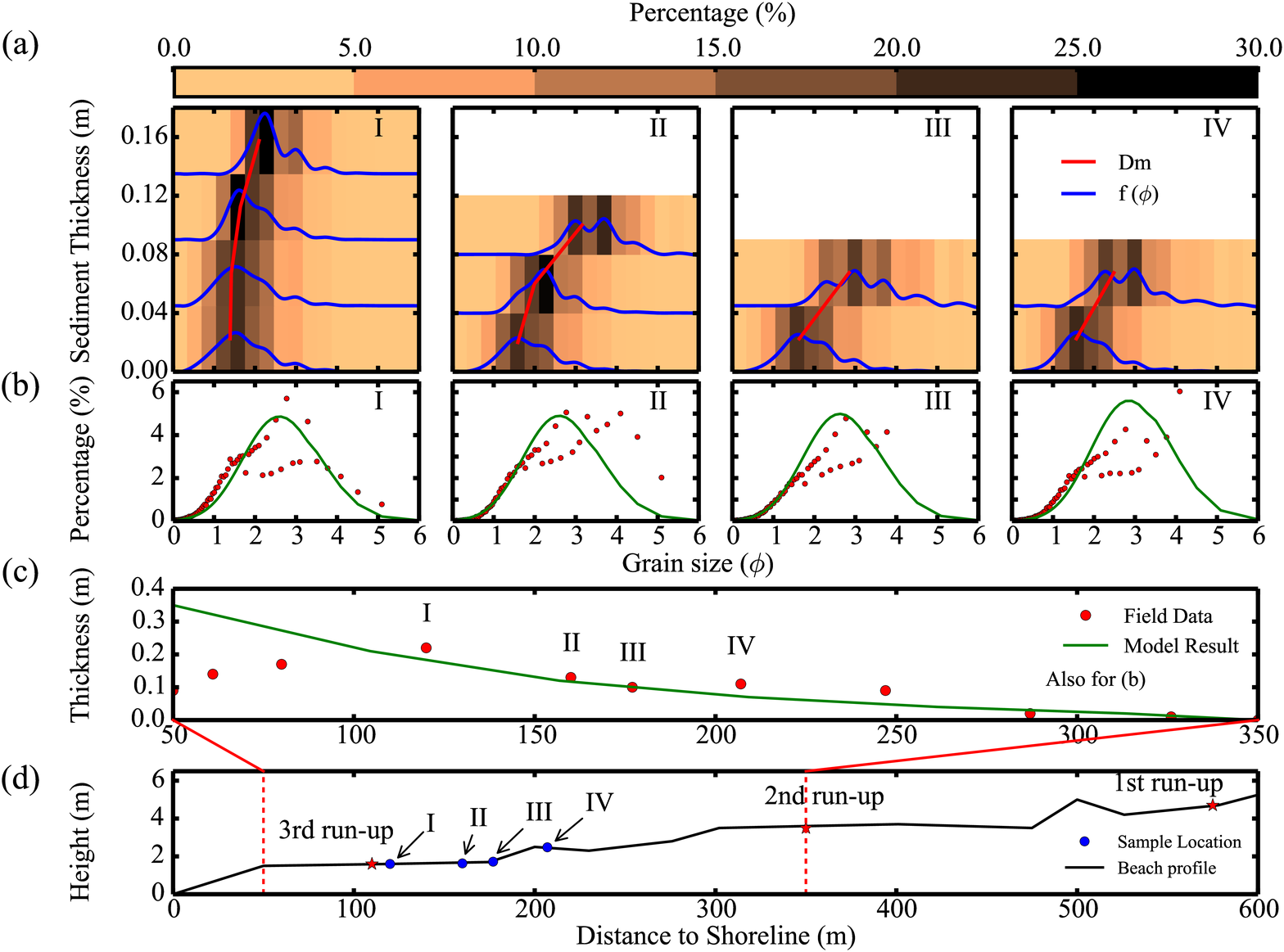}
  \caption{TSUFLIND simulation results and field measurement at Ranganathapuram, India. a: Vertical grading in grain size distribution (blue line) and mean grain size (red line) for four sampled locations (120m, 160m, 177m and 207m); b: the entire tsunami deposit grain-size distributions used as inputs to TSUFLIND (red points) and model result outputs from TSUFLIND (green line); c: tsunami deposit thickness field measurements (red points) and simulation results from TSUFLIND (green line); d: topography, wave run up and sample locations for test case ($I$ : 120m; $II$ : 160m; $III$: 177m; $IV$ : 207m).}
\end{figure}

\subsection{Field Observation and Data}
We employ the field data \citep{Bahlburg2007} from the 2004 Indian Ocean tsunami to demonstrate the capabilities of our framework (Figure 3). These samples come from the coastal area in the vicinity of Ranganathapuram, India. \citet{Bahlburg2007} identify sediment layers formed by the tsunami in this cross section and described grain-size distributions for each layer. There are some grass runners on the top of the tsunami sediment, which indicate the return flow direction and the erosion caused by the return flow. Most grain-size distributions of the sediment layers in the test case are unimodal (Figure 3b). Tsunami deposits in this cross section are usually well sorted, and the mean grain size is between 0.5 and 1.5 in $\phi$ scale, which corresponds to medium and coarse sand. Furthermore, \citet{Bahlburg2007} observe that the mean grain size is upward and landward fining in this cross section. Several sedimentary data, such as the deposit thickness and the grain-size distribution, will be used as input to TSUFLIND. The range of the grain sizes in the sediment source is based on all field samples collected during the field survey. Flow depth in this model will take full use of both the field observations and the model results from Soulsby’s model.

\subsection{Sedimentary Simulation Results}
TSUFLIND first simulates tsunami deposit thickness (Figure 3c). In the test case, the largest observed thickness is about 0.22 meters at 120 meters inland. For the first 100 meters in this cross section, the simulated thickness from TSUFLIND is obviously larger than the field measurement.  After 200 meters inland, the simulated thicknesses decrease quickly and generally fit with the field measurement.

TSUFLIND reconstructs sediment grain-size distributions for both the entire tsunami deposit and several vertical intervals at any given sample locations. The error of the entire tsunami sediment grain-size distribution in this test case is from 0.38\% to 1.54\%, which can be considered good simulation results. The error is less than 1.0\% from 120 meters to 160 meters inland and then increases to 1.5\% after 160 meters inland. We use four sediment samples to calculate grain-size distributions (Figure 3d $I-IV$, response to 120m, 160m, 177m and 207m from shoreline). Beyond 160 meters inland, there are fewer coarse grains and more fine grains in the simulated grain-size distribution than the field measurement (Figure 3d, $I$, $III$ and $IV$).

In order to study how the grain-size distribution changes in the vertical direction, we employ the new sediment formation module to simulate tsunami deposit grading. Figure 3a shows grain-size distribution for several vertical intervals at four different study locations. The grain size for these reconstruction results ranges from 0 to 6 in $\phi$ scale, so the portions of simulation results that are outside of this window are not plotted in Figure 3a. The number of vertical intervals separated decrease away from the shoreline as the sediment thickness decreases to zero. These simulation results show some features similar to the field observation, such as fining inland and upward. 

Based on the grain-size distribution for each vertical interval shown in Figure 3a, the sediment parameters used to describe the deposits can be calculated. Taking the sample from 120 meters inland as an example, we calculate the mean grain size, kurtosis, skewness and sorting factor for each interval. In the bottom several centimeters, the mean grain size does not change significantly and is about 1.2 in $\phi$ scale. Then the mean grain size decreases upward to 2.2 in $\phi$ scale. The change in kurtosis is about 0.8 to 1.1 in this sample, which indicates the grain-size distribution has a wider grain size range than a normal distribution. Sediment simulation results in this example also show that tsunami sediment changes from moderate sorting at the bottom to well sorting at the top where fine grains become dominant. Also the grain-size distribution is positively skewed which indicates that the distribution is skewed to fine grains.

\subsection{Hydrodynamic Inversion Results}

\begin{figure}[h!]
  \centering
    \includegraphics[width=1.0\textwidth]{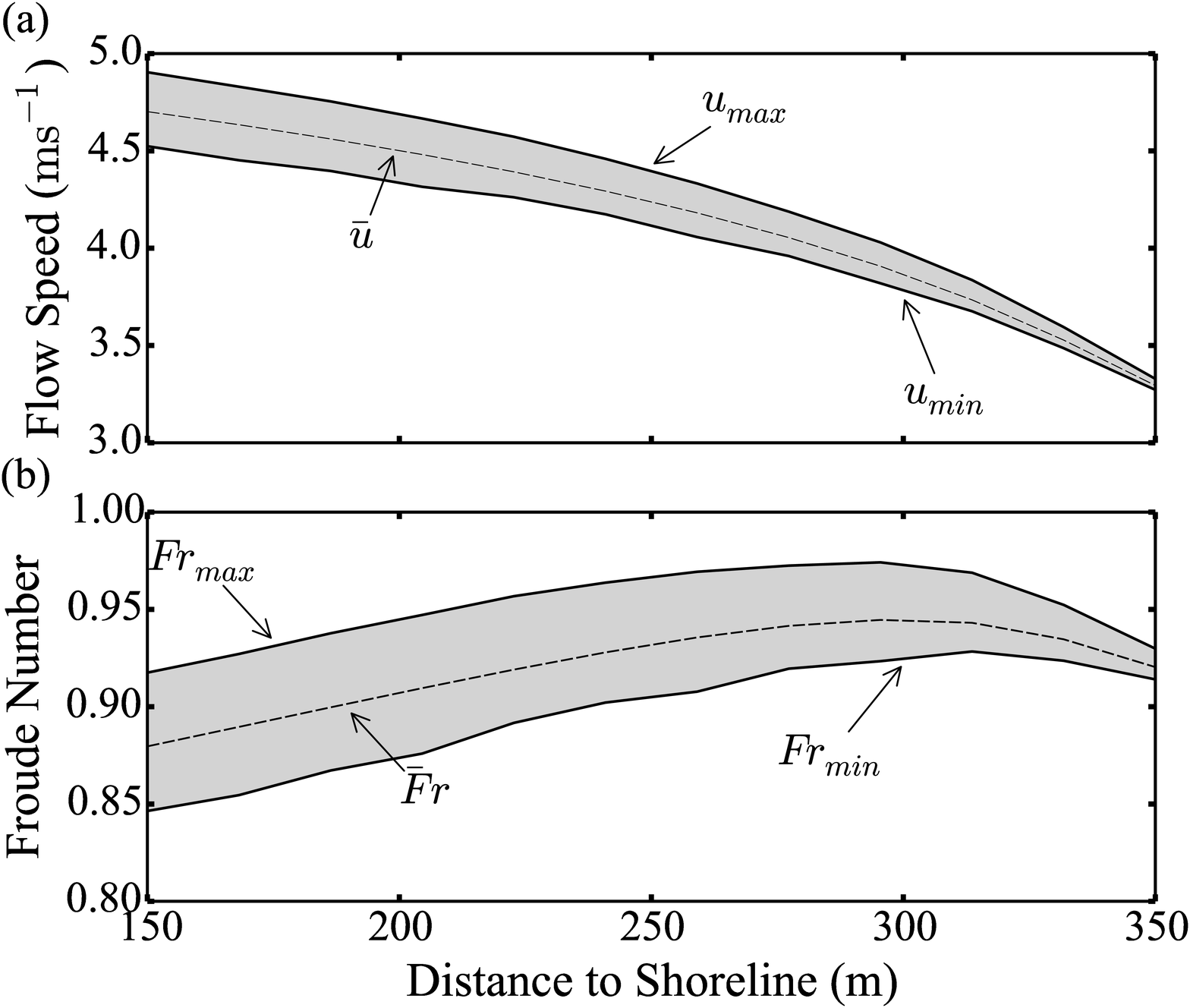}
  \caption{ The estimated flow speeds and Froude numbers from TSUFLIND. a: Tsunami flow speed estimates are indicated by the gray area with the boundaries of maximum and minimum possible speeds. The dashed line is the average value of estimated flow speeds. b: Froude number estimates are indicated by the gray area in this figure with the maximum and minimum possible values. The dashed line is the average value of Froude number. }
\end{figure}

After reconstructing grain-size distributions, TSUFLIND calculates the flow speed and Froude number at the sample locations. In the test case, Figure 4a and 4b show the flow speed and Froude number distribution along the slope.  The average flow speed decreases from 4.7 $\text{ms}^{-1}$ at 150 meters inland to 3.3 $\text{ms}^{-1}$ at 350 meters inland. The Froude number, which is around 0.9, does not change significantly in space. Also the range of possible velocities and Froude numbers decreases from 150 meters to 350 meters inland. 

The flow speed profile shape is influenced by the eddy viscosity profile and shear velocity. The eddy viscosity profile parameterizes the vertical distribution of turbulent stress. TSUFLIND follows the flow eddy viscosity profile based on laboratory data from \citet{gelfenbaum1986}. The flow speed gets the largest value on the water surface and decreases toward the sediment bed. TSUFLIND only gives the depth-averaged values as final results.

TSUFLIND computes the water surface profile to estimate the water volume when the tsunami wave reaches the maximum inundation. With the help of Eq (16), the wave amplitude can be estimated based on the slope ($m$) and the water volume ($V$). For the Eastern India case, the wave amplitudes range from 5 to 7 meters and the wavelength is close to 50 km.

\section{Discussion}
\label{S:4}
\subsection{Interpretation of test case results}
With the help of the presented model, we can reproduce tsunami sediments as well as infer the flow condition that generated them based on observations and analytical results of existing tsunami deposits. Figure 3 summarizes the results of our simulation for the tsunami deposits. The apparent difference of the deposit thickness between model results and observations for distances smaller than 120 m from the shoreline can be explained by strong return flow or large velocities from subsequent waves with small inundation. For distances from the coastline larger than 120 meters, the deposit thickness simulation results matched well with the observations. However, the observations are slightly larger due to the presence of topography change that may slow down the flow (Fig 3c and d). The finer grain-sizes contain the largest error between observation and model results is likely that the topography change is the main source of the error. However, the difference could also be a part of the model uncertainty. 

The calculated mean speed decreases from 4.7 $\text{ms}^{-1}$ to 3.3 $\text{ms}^{-1}$ along the studied section. The speed decreases continuously shown as Fig 4a, the Froude number increases and then decreases (Figure 4b). The mean Froude number is around 0.9 for this test case. As the change rate of water depth is a constant based on Sousby’s model, the mean Froude number changes only depends on the velocity changes.  At first, the flow speed decreases slower than the water depth, so the mean Froude number increases in this area (150 m to 300m). After some point around 300 meters, the flow decelerates quickly and causes the Froude number to decrease. The flow speed and Froude number results from TSUFLIND are shown as ranges of possible values with uncertainties (Figure 4a and 4b). In Figure 4, the gray area with solid line boundaries is the estimated flow speed or Froude number with maximum and minimum possible values. The dashed line is the mean value of the estimated flow speed or Froude number. The ranges of the speed and the Froude number decrease from 150m to 350m, which indicates the uncertainties decrease towards the sample location close to the landward sediment pinch-out. It is possible that the tsunami deposits near the maximum run-up position become thin, well-sorted and fine-grained containing less information about the flow condition. Tsunami wave amplitudes calculated by TSUFLIND are usually larger than real amplitudes, because the mathematical relationship (Eq 16) is from frictionless shallow water equations that ignore the effect of friction.

\subsection{Model limitation and improvement}
In this study, we combine three tsunami inversion models to simulate tsunami deposit and estimate tsunami flow parameters.  All three of these models are based on some model-specified basic assumptions. A significant assumption of TSUFLIND is that the sediment transport and the deposition process during a tsunami are considered  uniform in space and time. Consequently, the deposit comes from both horizontal convergence and suspension settling. TSUFLIND combines Sousby's model and TsuSedMod to simulate these two processes. This combination greatly improves the grain-size distribution simulation results. However, when the tsunami flow decelerates rapidly because of bathymetry changes or any other reasons, some part of the sediment deposited before will be eroded again. If the flow is strong enough, a significant part of tsunami deposit may be eroded, just like the result shown in Figure 3c from shoreline to 100 meters in land. As a result, the tsunami speed calculated by TSUFLIND will be underestimated. 

Another significant assumption of TSUFLIND is that most of tsunami deposits is transported by the suspension load and ignores the contribution of the bed load. This assumption will overestimate the tsunami flow speed and increase the percentage of coarse grains in the grain-size distribution. TSUFLIND is not applicable where bed load is the dominant sediment transport type. In most tsunami cases, tsunami deposits include both bed load and suspended load. In order to reduce the effect of bed load, only the suspension-grading part can be considered to estimate the flow speed. Suspension-grading in sediment is difficult to identify only based on the grain-size distribution. The simulated suspended-grading part has some difference with field observations, which also introduces uncertainties to the flow speed calculation. 

TSUFLIND has two aspects that require improvement: the applicable flow condition and the accuracy of final outputs including sediment simulation and flow speed calculation. The improvement of the applicable area can be made by employing other tsunami propagation models instead of Sousby's model, which can deal with non-uniform and unsteady flows. At the same time, the new model needs to consider both the suspension load and bed load. Also a new method for combining forward and inverse modeling will hold great promise for deciphering quantitative information from tsunami deposits and decreasing the uncertainties in tsunami sediment transport simulation and inversion results \citep{Sugawara2014295}.
\section{Conclusion}
\label{S:4}
Modeling the tsunami sediment deposition processes and estimating tsunami flow parameters will greatly improve not only the understanding of deposition from tsunami but also the risk assessment for extreme high-energy events. We presented a joint inversion model for tsunami deposit simulation and flow condition estimations in this paper.  

The combination of different inversion models allows the computation of a wide range of tsunami wave impacts or characteristics, ranging from sediment thickness, grain size distribution to flow speed and wave amplitude. However, these results are restricted by the flow condition. For instance the flow speed of non-uniform or unsteady flow cannot be inversed yet. Also if there were strong return flows or subsequent waves, the results from TSUFLIND will contain some uncertainties.  From a general point of view, with a simple bathymetry, the tsunami and paleotsunami as well as other extreme events can be understood with the help of this joint inversion framework.  More research needs to be done to improve this joint inversion framework presented to quantify and reduce the uncertainties in the inversion results and expand applicable conditions. 

\section{Appendix I}
\begin{footnotesize}
\begin{longtable}{ l l l }
\caption {Symbols List}\label{Tab:1}\\\toprule
\endfirsthead
\caption* {\textbf{Table \ref{Tab:1} Continued:} Symbols List}\\
\endhead
\endfoot
\endlastfoot
  Symbol    			&Unit					& Description																\\
  \hline 
  $h$ 					&m						&Water depth 																\\
  $w_s$ 				&$\text{ms}^{-1}$ 		&Settling velocity of the sediment grain									\\
  $l$ 					&m						&Horizontal distance a grain travels to be deposited  						\\
  $\gamma$				&1						&Run-up time constant														\\
  $T$ 					&s						&Inundation time 															\\
  $H$ 					&m 						&Maximum flow depth at a given location during tsunami inundation 			\\
  $\Delta h$ 			&m 						&Water depth increment due to tsunami 										\\
  $m$ 					&1 						&Slope 																		\\
  $R_z$ 				&m 						&Vertical water inundation limit 													\\
  $H_0$ 				&m 						&Maximum water depth at the shoreline 										\\
  $\zeta_{0}^{(i)}$ 	&m					    &Thickness of the deposit for grain size $i$ at the shoreline 				\\
  $T_d$ 				&s				 		&Deposition time 															\\
  $C_{0}^{(i)}$ 		&$\text{m}^3/\text{m}^3$&Depth averaged sediment concentration for grain size $i$ 					\\
  $p$ 					&1		 				&Porosity 																	\\
  $\zeta^{(i)}$ 		&m	 					&Sediment thickness for grain size $i$ 										\\
  $R_{s}^{(i)}$ 		&m 						&Distance between sediment extend and the shoreline for grain size $i$ 		\\
  $\Delta \eta^{(i)}$ 	&m						&Sediment thickness increment for grain size $i$											\\
  $C^{(i)}(z)$ 			&$\text{m}^3/\text{m}^3$&Sediment concentration profile for grain size $i$ 							\\
  $z_o$ 				&m						&Bottom roughness 															\\
  $K(z)$ 				&$\text{kg}\text{m}^{-1}\text{s}^{-1}$&Eddy viscosity profile										\\
  $C_{r}^{(i)}$ 		&$\text{m}^3/\text{m}^3$&Reference sediment concentration 											\\
  $\beta_{0}$ 			&1 						&Resuspension coefficient 													\\
  $f^{(i)}$ 			&\%						&Percentage of the sediment of grain size $i$ 								\\
  $S^{(i)}$ 			&1 						&Normalized excess shear stress 											\\
  $\tau_{b}$ 			&$\text{kg}\text{m}^{-1}\text{s}^{-2}$&Bed shear stress 											\\
  $\tau_{cr}^{(i)}$ 	&$\text{kg}\text{m}^{-1}\text{s}^{-2}$&Critical shear stress of the initial sediment movement for grain size $i$  \\
  $N$ 					&1 						&Number of grain size classes 												\\
  $z_{j}$ 				&m 						&Sediment original elevation 												\\
  $t^{(i)}_{j}$ 		&s 						&Deposition time for grain size $i$ sediment at elevation $z_{j}$ 			\\
  $L_{k}$ 				&1 						&Second norm of error for layer $k$'s grain-size distribution 				\\
  $f^{(i)}_{m}$ 		&\%						&Modeled percentages for grain size class $i$ 							\\
  $f^{(i)}_{o}$ 		&\%						&Observed percentages for grain size class $i$ 						\\
  $\overline{L}$ 		&1 						&Average second norm value of grain-size distribution for a location				\\
  $L_{th}$ 				&1 						&Second Norm value of thickness between the model result and the field observation \\
  $th_{m}$ 				&m 						&Modeled thicknesses   													\\
  $th_{f}$ 				&m 						&Observed thicknesses 												\\
  $Q$ 					&1 						&Number of sample locations													\\
  $\xi$ 				&m 						&Offshore wave amplitude													\\
  $V$ 					&$\text{m}^3$			&Water volume that covers the land at maximum inundation 					\\
  $R_w$ 				&m					    &Distance of water run-up to shoreline											\\
  $\eta(x,t)$ 			&m 						&Solitary wave form															\\
  $C$ 					&$\text{ms}^{-1}$ 		&Wave celerity																\\
  $u(x,t)$ 				&$\text{ms}^{-1}$  		&Horizontal velocity in shallow water equations								\\
  $g$ 					&$\text{ms}^{-2}$ 		&Gravitational constant														\\
  $B(x)$ 				&m 						&Bed topography function 													\\
  $d$					&m						&Water depth of continental shelf \\
  \hline
\end{longtable}
\end{footnotesize}

\section{Appendix II}
TSUFLIND implements a simplified method to estimate the representative offshore tsunami wave amplitude. First of all, the water volume on the land due to tsunami wave is calculated by:

\begin{equation}
\label{eq:17}
V=\int_{0}^{R_w}{h(x)dx}
\end{equation}
where $V$ is the water volume, $R_w$ is the distance of run-up to shoreline, $h(x)$ is water depth distribution function on land. To simplify this problem, we assume tsunami wave is a solitary wave. The solitary wave form is given as a function of distance $x$ and time $t$ by 

\begin{equation}
\label{eq:18}
\eta(x,t)=\xi sech^2(k(x-Ct))
\end{equation}
where 

\begin{equation}
\label{eq:19}
k=\sqrt{\frac{3\xi}{4d^3}}
\end{equation}
$\xi$ is the wave amplitude and $d$ is the water depth of continental shelf, which is assumed as 500 meters. $C $ is the wave celerity which is expressed as:

\begin{equation}
\label{eq:20}
C=\sqrt{g(h+H)}
\end{equation}
Initial velocity in shallow water equation code is set as:

\begin{equation}
\label{eq:21}
v_0(x,t)=\sqrt{g/h_0}\eta_{(x,t=0)}
\end{equation}
We calculate the water volume when tsunami wave got the maximum run-up based on water distribution function $h(x)$. Water depth function $h(x)$ comes from a shallow water equations code. The shallow water equations code used here originally is designed for studying the propagation and runup of solitary wave by using a high-resolution finite volume method to solve following equations\citep{FLD:FLD1537}:

\begin{equation}
\label{eq:22}
\frac{\partial h}{\partial t}+\frac{\partial (uh)}{\partial x}=0
\end{equation}

\begin{equation}
\label{eq:23}
\frac{\partial (uh)}{\partial t}+\frac{\partial}{\partial x}(hu^2+\frac{1}{2}gh^2)=-gh\frac{dB}{dx}
\end{equation}
where $h(x,t)$ is the flow depth, $u(x,t)$ is the horizontal velocity, $g$ is the gravitational constant, $B(x)$ is the bed topography function. 

In this code, a conservative form of the nonlinear shallow water equations with source term is solved by using a high-resolution Godunov-type explicit scheme with Roe’s approximate Riemamn solver \citep{FLD:FLD1537}. In order to get the mathematical relationship between the maximum water volume ($V$), slope ($m$) and initial wave amplitude ($\xi$), we design a parameter study by varying slope and wave amplitude to calculate the water volume. Finally, we use curve fitting methods to get the mathematical relationship based on parameter study data set.   

\bibliographystyle{apa}
\bibliography{TSUFLIND}

\begin{thebibliography}{}

\bibitem[\protect\astroncite{Bahlburg and Weiss}{2007}]{Bahlburg2007}
Bahlburg, H. and Weiss, R. (2007).
\newblock Sedimentology of the december 26, 2004, sumatra tsunami deposits in
  eastern india (tamil nadu) and kenya.
\newblock {\em International Journal of Earth Sciences}, 96(6).

\bibitem[\protect\astroncite{{Benner} et~al.}{2010}]{benner2010}
{Benner}, R., {Browne}, T., {Br\"u{}ckner}, H., {Kelletat}, D., and
  {Scheffers}, A. (2010).
\newblock Boulder transport by waves: Progress in physical modelling.
\newblock {\em Zeitschrift für Geomorphologie, Supplementbände},
  54(3):127--146.

\bibitem[\protect\astroncite{Bourgeois et~al.}{1988}]{Bourgeois1988}
Bourgeois, J., Hansen, T.~A., Wiberg, P.~L., and Kauffman, E.~G. (1988).
\newblock A tsunami deposit at the cretaceous-tertiary boundary in texas.
\newblock {\em Science}, 241(4865):pp. 567--570.

\bibitem[\protect\astroncite{Delis et~al.}{2008}]{FLD:FLD1537}
Delis, A.~I., Kazolea, M., and Kampanis, N.~A. (2008).
\newblock A robust high-resolution finite volume scheme for the simulation of
  long waves over complex domains.
\newblock {\em International Journal for Numerical Methods in Fluids},
  56(4):419--452.

\bibitem[\protect\astroncite{Gelfenbaum and Smith}{1986}]{gelfenbaum1986}
Gelfenbaum, G. and Smith, J.~D. (1986).
\newblock Experimental evaluation of a generalized suspended-sediment transport
  theory.

\bibitem[\protect\astroncite{Hutchinson et~al.}{1997}]{hutchinson1997}
Hutchinson, I., Clague, J.~J., and Mathewes, R.~W. (1997).
\newblock Reconstructing the tsunami record on an emerging coast: A case study
  of kanim lake, vancouver island, british columbia, canada.
\newblock {\em Journal of Coastal Research}, 13(2):pp. 545--553.

\bibitem[\protect\astroncite{Jaffe et~al.}{2011}]{Jaffe201123}
Jaffe, B., Buckley, M., Richmond, B., Strotz, L., Etienne, S., Clark, K., Watt,
  S., Gelfenbaum, G., and Goff, J. (2011).
\newblock Flow speed estimated by inverse modeling of sandy sediment deposited
  by the 29 september 2009 tsunami near satitoa, east upolu, samoa.
\newblock {\em Earth-Science Reviews}, 107(1–2):23 -- 37.
\newblock The 2009 South Pacific tsunami.

\bibitem[\protect\astroncite{Jaffe and Gelfenbuam}{2007}]{Jaffe2007347}
Jaffe, B.~E. and Gelfenbuam, G. (2007).
\newblock A simple model for calculating tsunami flow speed from tsunami
  deposits.
\newblock {\em Sedimentary Geology}, 200(3–4):347 -- 361.
\newblock Sedimentary Features of Tsunami Deposits - Their Origin, Recognition
  and Discrimination: An Introduction.

\bibitem[\protect\astroncite{Jaffe et~al.}{2012}]{Jaffe201290}
Jaffe, B.~E., Goto, K., Sugawara, D., Richmond, B.~M., Fujino, S., and
  Nishimura, Y. (2012).
\newblock Flow speed estimated by inverse modeling of sandy tsunami deposits:
  results from the 11 march 2011 tsunami on the coastal plain near the sendai
  airport, honshu, japan.
\newblock {\em Sedimentary Geology}, 282(0):90 -- 109.
\newblock The 2011 Tohoku-oki tsunami.

\bibitem[\protect\astroncite{MacWilliams}{2004}]{MacWilliams2005}
MacWilliams, M.~L. (2004).
\newblock {\em Three-dimensional hydrodynamic simulation of river channels and
  floodplains}.
\newblock PhD thesis, Stanford University.

\bibitem[\protect\astroncite{Madsen et~al.}{1993}]{Madsen19931303}
Madsen, O., Wright, L., Boon, J., and Chisholm, T. (1993).
\newblock Wind stress, bed roughness and sediment suspension on the inner shelf
  during an extreme storm event.
\newblock {\em Continental Shelf Research}, 13(11):1303 -- 1324.

\bibitem[\protect\astroncite{Martin et~al.}{2008}]{martin2008}
Martin, M.~E., Weiss, R., Bourgeois, J., Pinegina, T.~K., Houston, H., and
  Titov, V.~V. (2008).
\newblock Combining constraints from tsunami modeling and sedimentology to
  untangle the 1969 ozernoi and 1971 kamchatskii tsunamis.
\newblock {\em Geophysical Research Letters}, 35(1):n/a--n/a.

\bibitem[\protect\astroncite{Moore et~al.}{2007}]{Moore2007336}
Moore, A.~L., McAdoo, B.~G., and Ruffman, A. (2007).
\newblock Landward fining from multiple sources in a sand sheet deposited by
  the 1929 grand banks tsunami, newfoundland.
\newblock {\em Sedimentary Geology}, 200(3–4):336 -- 346.
\newblock Sedimentary Features of Tsunami Deposits - Their Origin, Recognition
  and Discrimination: An Introduction.

\bibitem[\protect\astroncite{Nandasena and Tanaka}{2013}]{Nandasena2013163}
Nandasena, N. and Tanaka, N. (2013).
\newblock Boulder transport by high energy: Numerical model-fitting
  experimental observations.
\newblock {\em Ocean Engineering}, 57(0):163 -- 179.

\bibitem[\protect\astroncite{Noormets et~al.}{2004}]{Noormets200441}
Noormets, R., Crook, K.~A., and Felton, E.~A. (2004).
\newblock Sedimentology of rocky shorelines: 3.: Hydrodynamics of megaclast
  emplacement and transport on a shore platform, oahu, hawaii.
\newblock {\em Sedimentary Geology}, 172(1–2):41 -- 65.

\bibitem[\protect\astroncite{Nott}{1997}]{Nott1997193}
Nott, J. (1997).
\newblock Extremely high-energy wave deposits inside the great barrier reef,
  australia: determining the cause—tsunami or tropical cyclone.
\newblock {\em Marine Geology}, 141(1–4):193 -- 207.

\bibitem[\protect\astroncite{Smith et~al.}{2007}]{Smith2007362}
Smith, D., Foster, I., Long, D., and Shi, S. (2007).
\newblock Reconstructing the pattern and depth of flow onshore in a
  palaeotsunami from associated deposits.
\newblock {\em Sedimentary Geology}, 200(3–4):362 -- 371.
\newblock Sedimentary Features of Tsunami Deposits - Their Origin, Recognition
  and Discrimination: An Introduction.

\bibitem[\protect\astroncite{Soulsby et~al.}{2007}]{sousby2007}
Soulsby, R., Smith, D., and Ruffman, A. (2007).
\newblock {\em Reconstructing Tsunami Run-Up from Sedimentary Characteristics:
  A Simple Mathematical Model}, chapter~83, pages 1075--1088.

\bibitem[\protect\astroncite{Spiske et~al.}{2013}]{Spiske201331}
Spiske, M., Piepenbreier, J., Benavente, C., Kunz, A., Bahlburg, H., and
  Steffahn, J. (2013).
\newblock Historical tsunami deposits in peru: Sedimentology, inverse modeling
  and optically stimulated luminescence dating.
\newblock {\em Quaternary International}, 305(0):31 -- 44.
\newblock Ranked habitats and the process of human colonization of South
  America.

\bibitem[\protect\astroncite{Spiske et~al.}{2010}]{Spiske201029}
Spiske, M., Weiss, R., Bahlburg, H., Roskosch, J., and Amijaya, H. (2010).
\newblock The tsusedmod inversion model applied to the deposits of the 2004
  sumatra and 2006 java tsunami and implications for estimating flow parameters
  of palaeo-tsunami.
\newblock {\em Sedimentary Geology}, 224(1–4):29 -- 37.

\bibitem[\protect\astroncite{Sugawara et~al.}{2014}]{Sugawara2014295}
Sugawara, D., Goto, K., and Jaffe, B.~E. (2014).
\newblock Numerical models of tsunami sediment transport — current
  understanding and future directions.
\newblock {\em Marine Geology}, 352(0):295 -- 320.
\newblock 50th Anniversary Special Issue.

\bibitem[\protect\astroncite{Tang and Weiss}{2014}]{HUI2014}
Tang, H. and Weiss, R. (2014).
\newblock Tsuspeedv0.5: Inversion of flow depth and flow speed along a cross
  section.
\newblock In {\em 2014 AGU Fall Meeting}.

\bibitem[\protect\astroncite{Witter et~al.}{2012}]{witter2012}
Witter, R., Jaffe, B., Zhang, Y., and Priest, G. (2012).
\newblock Reconstructing hydrodynamic flow parameters of the 1700 tsunami at
  cannon beach, oregon, usa.
\newblock {\em Natural Hazards}, 63(1):223--240.

\end{thebibliography}

\end{document}